\documentclass[allclo]{FBSart}
\usepackage{amsmath}
\usepackage{amsfonts}
\usepackage{graphicx}
\usepackage{amssymb}

\begin{document}
\title{Light nuclei from chiral EFT interactions}
\author{P.~Navr\'atil$^1$\thanks{\textit{E-mail address:}navratil1@llnl.gov},
V.~G.~Gueorguiev$^1$, J.~P.~Vary$^{1,2}$, W.~E.~Ormand$^1$, A.~Nogga$^3$,
S.~Quaglioni$^1$}
\institute{$^1$Lawrence Livermore National Laboratory, L-414, P.O. Box 808,
             Livermore, CA  94551, USA\\
$^2$Department of Physics and Astronomy, Iowa State University, Ames,
Iowa 50011, USA\\
$^3$Forschungszentrum J\"{u}lich, Institut f\"{u}r Kernphysik (Theorie),
D-52425 J\"{u}lich, Germany}

\runningauthor{P.\,Navr\'atil, V.~G.~Gueorguiev, J.~P.~Vary, W.~E.~Ormand, 
A.~Nogga and S.~Quaglioni} 
\runningtitle{Light nuclei from chiral EFT interactions}
\sloppy

\maketitle
\begin{abstract}
Recent developments in nuclear theory allow us to make a connection
between quantum chromodynamics (QCD) and low-energy nuclear physics. 
First, chiral effective field theory ($\chi$EFT) 
provides a natural hierarchy to define two-nucleon (NN), 
three-nucleon (NNN), and even four-nucleon interactions. 
Second, {\em ab initio} methods have been developed 
capable to test these interactions for light nuclei. In this contribution, 
we discuss {\em ab initio}
no-core shell model (NCSM) calculations for $s$-shell and $p$-shell nuclei
with NN and NNN interactions derived within $\chi$EFT.
\end{abstract}
%
\section{\label{sec:intro}Introduction}

The major outstanding problem in nuclear physics is to calculate properties 
of finite nuclei starting from the basic interactions among nucleons. 
There are two aspects to this problem. First, the basic interactions among nucleons 
are complicated, they are not uniquely defined and there
is evidence that more than just two-nucleon forces are important. 
Second, the nuclear many-body problem is very difficult to solve. 
This is a direct consequence of the complex nature
of the inter-nucleon interactions. 

Interactions among nucleons are governed by QCD.
In the low-energy regime relevant to nuclear structure,
QCD is non-perturbative, and, therefore, hard to solve. Thus, theory has
been forced to resort to models for the interaction. 
New theoretical developments, however, allow us connect QCD
with low-energy nuclear physics through promising bridge of $\chi$EFT~\cite{Weinberg}.

\section{{\it Ab initio} no-core shell model}

In the {\it ab initio} NCSM, we consider a system 
of $A$ point-like non-relativistic nucleons that interact by realistic 
NN or NN+NNN interactions. Unlike in standard shell model calculations,
in the NCSM there is no inert core, all the nucleons are considered active. 
Therefore the ``no-core'' in the name of the approach.
Besides the employment of realistic NN or NN+NNN interactions, two other major features 
characterize the NCSM: i) the use of an harmonic oscillator (HO) basis 
truncated by a chosen maximal HO excitation energy $N_{\rm max}\hbar\Omega$ relative 
to the unperturbed ground state of the $A$-nucleon system; 
ii) the use of effective interactions.
The reason behind the choice of the HO basis is the fact that
this is the only basis that allows to use single-nucleon coordinates and consequently 
the second-quantization representation without violating the translational invariance
of the system. The powerful techniques based on the second quantization and developed 
for standard shell model calculations can then be utilized. Therefore the ``shell model''
in the name of the approach. As a downside, one has to face the consequences of the incorrect
asymptotic behavior of the HO basis. 
The second feature comes as a consequence of the basis truncation.
In order to speed up convergence with the basis enlargement, we construct an effective interaction
from the original realistic NN or NN+NNN potentials by means of a unitary transformation.
The effective interaction depends on the basis truncation and by construction recovers
the original realistic NN or NN+NNN interaction as the size of the basis approaches infinity.
In principle, one can also perform calculations with the unmodified, ``bare'', original 
interactions. Such calculations are then variational with respect to the HO frequency $\Omega$ 
and the basis truncation parameter $N_{\rm max}$.

\section{Light nuclei from chiral EFT interactions}

Currently the most promising approach to the construction of accurate inter-nucleon 
forces from QCD is the $\chi$EFT.
The $\chi$EFT predicts, along with the NN interaction
at the leading order, an NNN interaction starting at the 3rd order 
(next-to-next-to-leading
order or N$^2$LO)~\cite{Weinberg,vanKolck:1994}, 
and even an NNNN interaction starting at the 4th order (N$^3$LO)~\cite{Epelbaum06}.
The details of QCD dynamics are contained in parameters,
low-energy constants (LECs), not fixed by the symmetry. These parameters
can be constrained by experiment. 
A crucial feature of $\chi$EFT is the consistency between
the NN, NNN and NNNN parts. As a consequence, at N$^2$LO and N$^3$LO, except
for two LECs, assigned to two NNN diagrams,
the potential is fully
constrained by the parameters defining the NN interaction.

\begin{figure}[hbtp]
\begin{minipage}{65mm}
  \includegraphics*[width=0.95\columnwidth]
   {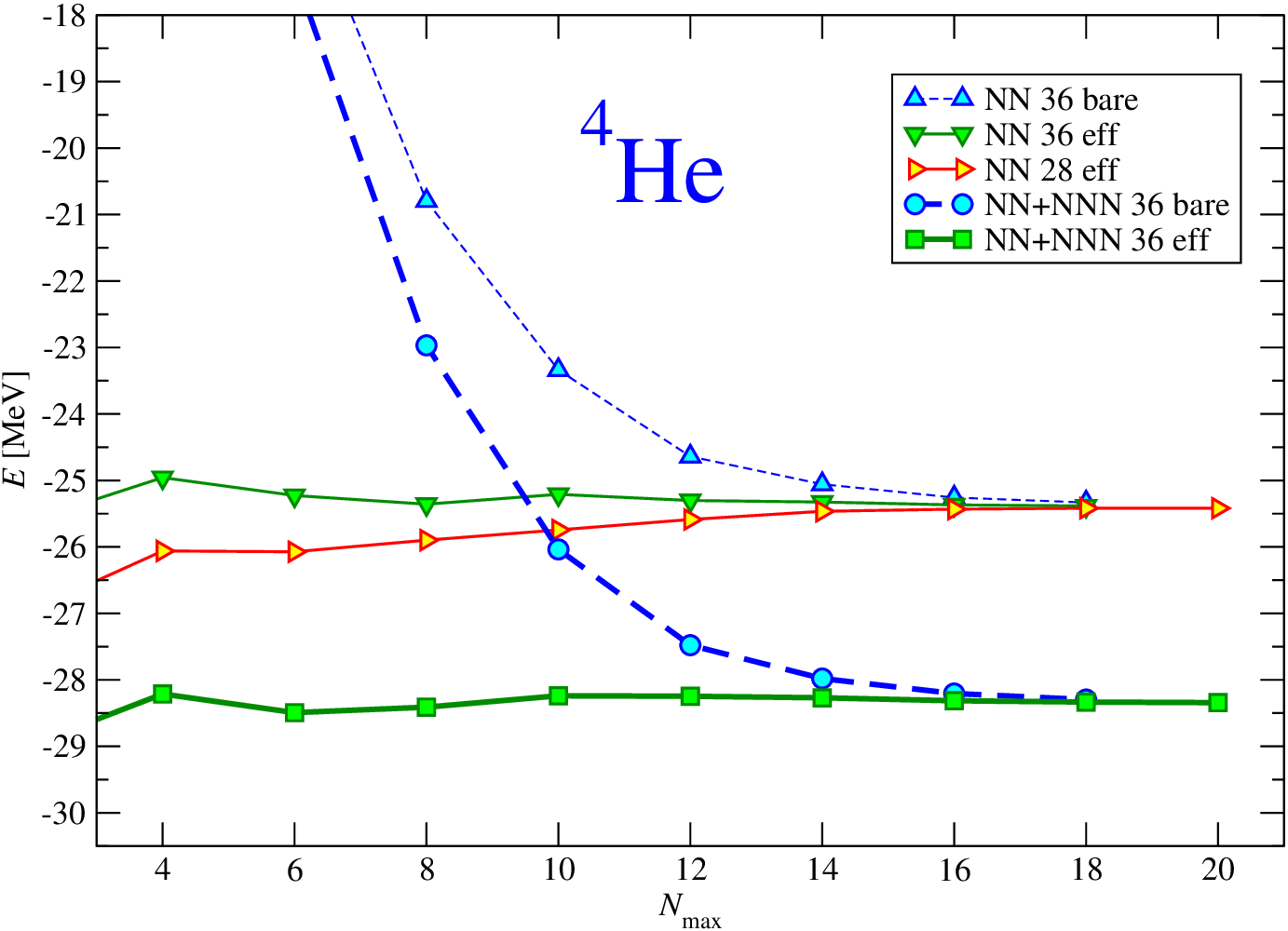}
  \caption{$^4$He ground-state energy dependence on the size of the basis.
The HO frequencies of $\hbar\Omega=28$ and 36~MeV were employed. Results with (thick lines)
and without (thin lines) the NNN interaction are shown. The full lines correspond 
to calculations with three-body effective interaction, 
the dashed lines to calculations with the bare interaction. 
  \label{gs_He4}}
\end{minipage}
\hfill
\begin{minipage}{65mm}
{\includegraphics[width=0.9\columnwidth]{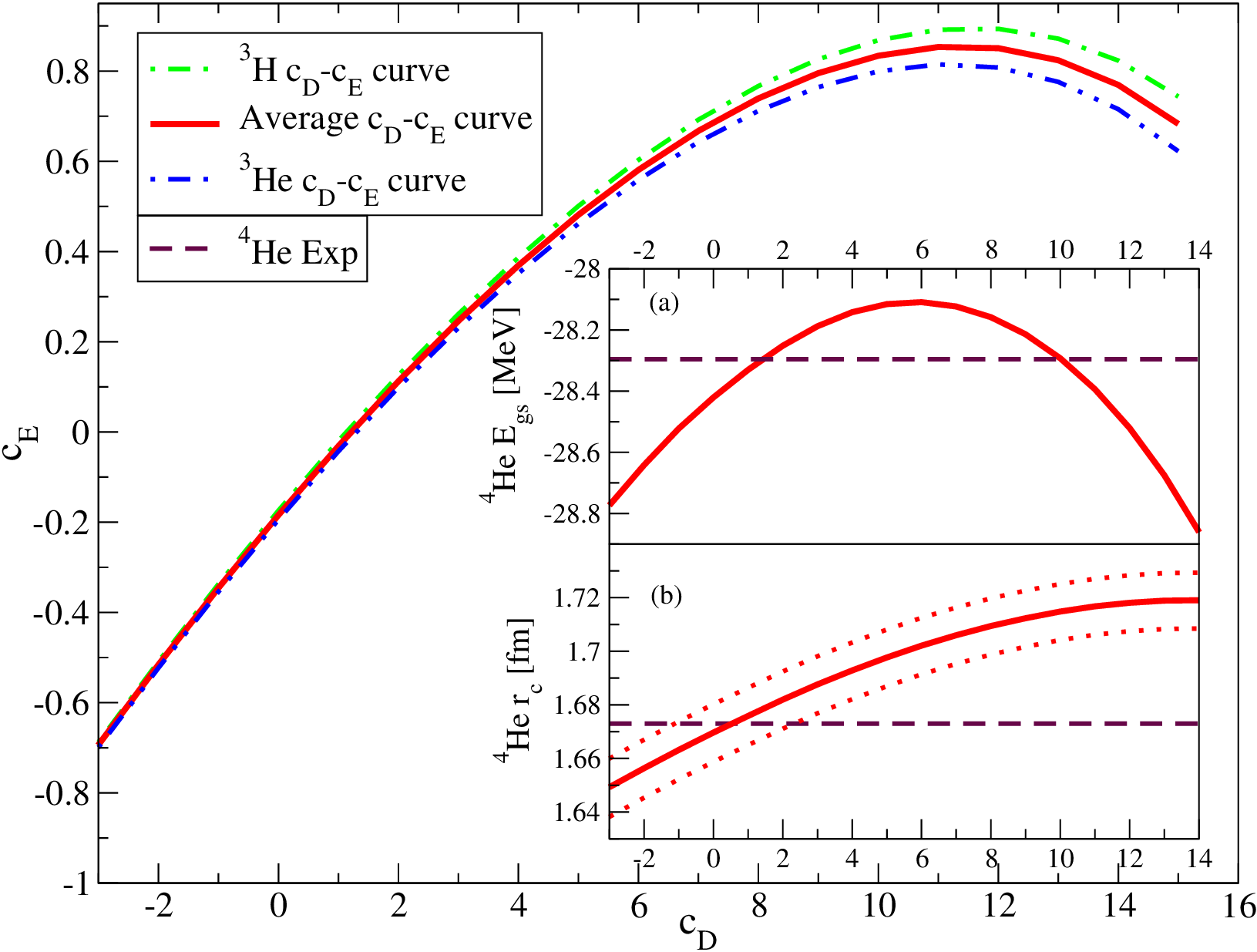}}
\caption{Relations between $c_D$ and $c_E$ for which  the 
binding energy of $^3$H ($8.482$ MeV) and  $^3$He ($7.718$ MeV) are reproduced. 
(a) $^4$He ground-state energy along the averaged curve. 
(b) $^4$He charge radius $r_c$ along the averaged curve. Dotted lines represent 
the $r_c$ uncertainty due to the uncertainties in the proton charge radius.}
\label{CDCE_curve}
\end{minipage}
\end{figure}

We adopt the potentials of the $\chi$EFT at the orders presently available, 
the NN at N$^3$LO of Ref.~\cite{N3LO} and the NNN interaction 
at N$^2$LO \cite{vanKolck:1994}. 
We use {\it ab initio} NCSM calculations in two
ways. One of them is the determination of the LECs assigned to two NNN diagrams,
$c_D$ and $c_E$ \cite{Nogga06} that must be determined in $A\geq 3$ systems.
$c_D$ ($c_E$) is the strength of the $NN$-$\pi$-$N$ ($NNN$) contact term.
The other is testing predictions 
of the chiral NN and NNN interactions for light nuclei. 

It is important to note that our NCSM results through $A=4$ are fully converged in that 
they are independent of the $N_{\rm max}$ cutoff and the $\hbar\Omega$ HO energy.
This is demonstrated in Fig.~\ref{gs_He4}, where convergence of the $^4$He ground-state
energy using $\chi$EFT interactions with and without the NNN terms is shown.
For heavier systems, we characterize the approach to convergence by the dependence 
of results on $N_{\rm max}$ and $\hbar\Omega$.

Fig.~\ref{CDCE_curve} shows the trajectories of the two LECs $c_D-c_E$ that 
are determined from fitting the binding energies of the $A=3$ systems. 
Separate curves are shown for $^3$H and $^3$He fits, as well as their average.
There are two points where the binding of $^4$He is reproduced exactly.  
We observe, however, that in the whole investigated range of $c_D-c_E$, the calculated 
$^4$He  binding energy  is within a few hundred keV of experiment. 
Consequently, the determination of the LECs in this way is likely not very stringent.
We therefore investigate the sensitivity of the $p$-shell nuclear properties 
to the choice of the $c_D-c_E$ LECs. First, we maintain the $A=3$ binding energy 
constraint. Second, we limit ourselves to the $c_D$ values in the vicinity 
of the point $c_{D}\sim 1$ since the values close to the point 
$c_{D}\sim 10$ overestimate the $^4$He radius.

\begin{figure}
\begin{minipage}{65mm}
{\includegraphics[width=0.9\columnwidth]{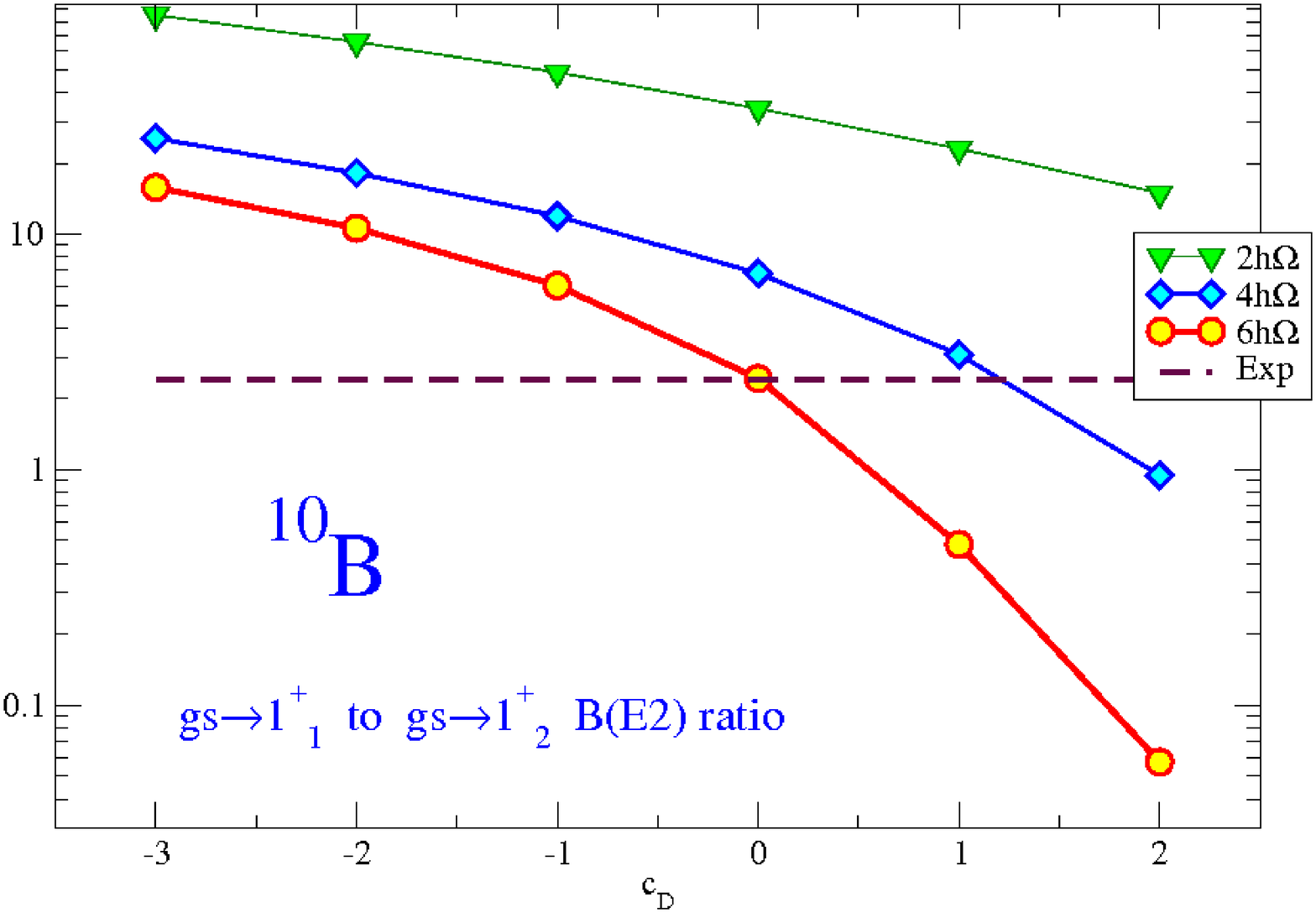}}
\caption{Dependence of the $^{10}$B 
B(E2;$3^+_1 0 \rightarrow 1^+_1 0$)/B(E2;$3^+_1 0 \rightarrow 1^+_2 0$) ratio
on $c_D$, with $c_E$ constrained by the $A=3$ binding energy fit, 
for different basis sizes.
The HO frequency of $\hbar\Omega=14$ MeV was employed.}
\label{cD_dep_10B}
\end{minipage}
\hfill
\begin{minipage}{65mm}
{\includegraphics[width=0.88\columnwidth]{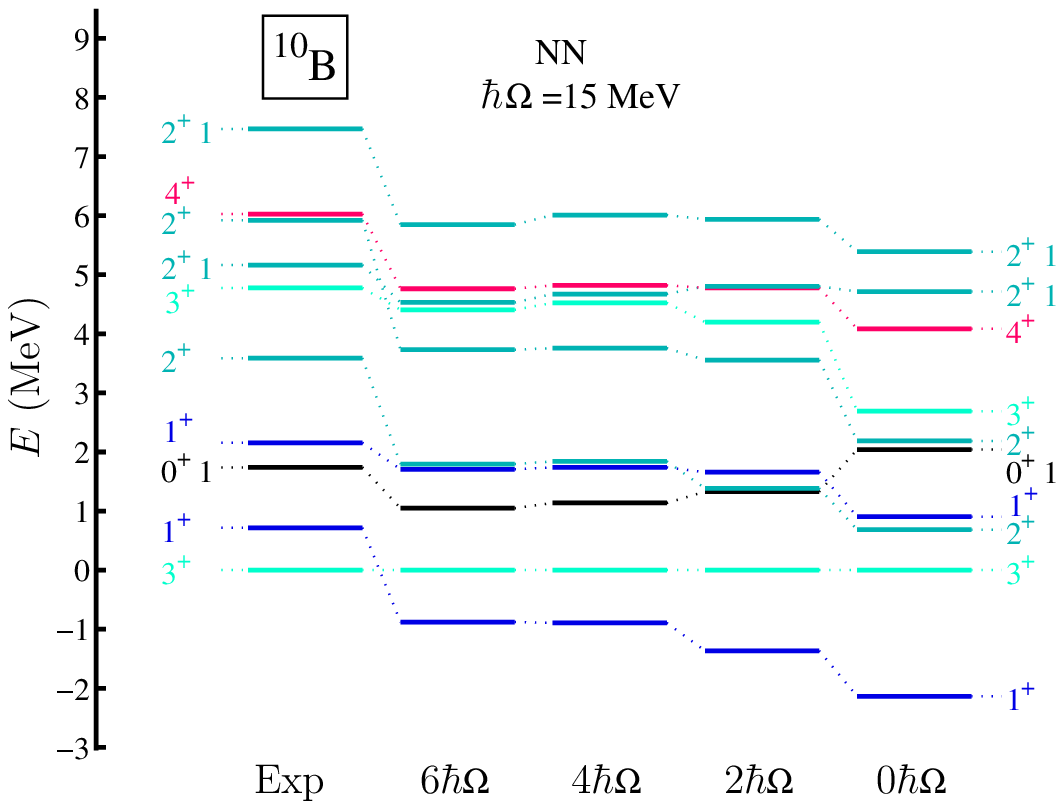}}
\caption{$^{10}$B excitation spectra as function of the basis space size $N_{\rm max}$ 
at $\hbar\Omega=15$ MeV using the chiral NN interaction
compared to experiment. The isospin of the states 
not explicitly depicted is $T$=0.}
\label{B10_NN}
\end{minipage}
\end{figure}

While most of $p$-shell nuclear properties, e.g. excitation spectra,
are not very sensitive to variations of $c_D$ in the vicinity of the $c_{D}\sim 1$ point,
we were able to identify several observables that do demonstrate strong dependence on $c_D$. 
For example, the $^6$Li quadrupole moment that changes sign 
depending on the choice of $c_D$. In Fig.~\ref{cD_dep_10B}, we display the ratio 
of the B(E2) transitions from the $^{10}$B ground state to the first and the second $1^+ 0$ state.
This ratio changes by several orders of magnitude 
depending on the $c_D$ variation. This is due to the fact
that the structure of the two $1^+ 0$ states is exchanged depending on $c_D$. 
From Figs.~\ref{CDCE_curve} and \ref{cD_dep_10B}, we can see that for $c_D<-2$ the $^4$He radius
underestimate experiment while for $c_D>0$ the lowest
two $1^+$ states of $^{10}$B are reversed.
We therefore select $c_D=-1$ for our further investigation.

We present in Fig.~\ref{B10_NN} the excitation spectra of $^{10}$B as 
a function of $N_{\rm max}$ for the chiral NN interaction alone. The convergence 
with increasing $N_{\rm max}$ is quite reasonable for the low-lying states. 
Similar convergence rates are obtained for other $p-$shell nuclei that we investigate.
A remarkable feature of the $^{10}$B results is the observation that the chiral NN
interaction alone predicts incorrect ground-state spin of $^{10}$B. The experimental
value is $3^+ 0$, while the calculated one is $1^+ 0$. On the other hand, once we also 
include the chiral NNN interaction in the Hamiltonian, which is actually required by 
the $\chi$EFT, the correct ground-state spin is predicted. Further, once we select the $c_D$
value as discussed above, i.e. $c_D=-1$, we also obtain the two lowest $1^+ 0$ states
in the experimental order.  

\begin{figure}[htb]
{\includegraphics[width=1.0\columnwidth]{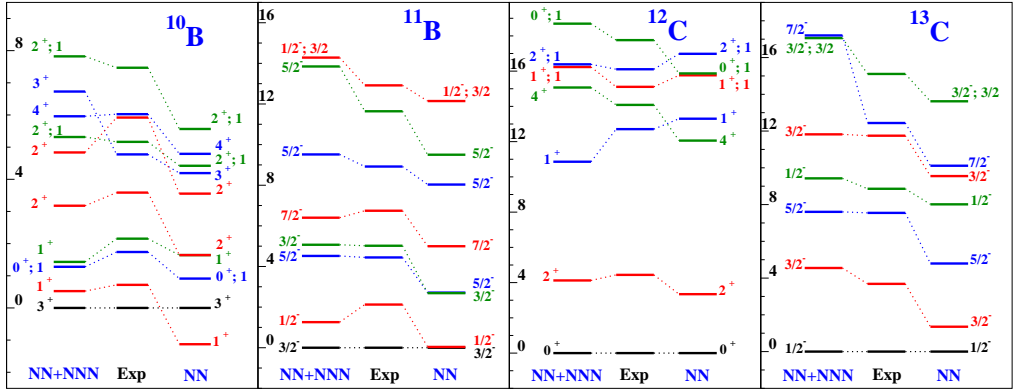}}
\caption{States dominated by $p$-shell configurations for $^{10}$B, $^{11}$B, $^{12}$C, 
and $^{13}$C calculated at $N_{\rm max}=6$ using  $\hbar\Omega=15$ MeV 
(14 MeV for $^{10}$B). Most of the eigenstates are isospin $T$=0 or 1/2, 
the isospin label is explicitly shown only for states with $T$=1 or 3/2.
The excitation energy scales are in MeV.}
\label{B10B11C12C13}
\end{figure}

We display in Fig.~\ref{B10B11C12C13} the natural parity excitation spectra of four nuclei 
in the middle of the $p-$shell with both the NN and the NN+NNN interactions 
from $\chi$EFT. The results shown are obtained 
in the largest basis spaces achieved to date for these nuclei with the NNN interactions, 
$N_{\rm max}=6$ ($6\hbar\Omega$). Overall, the NNN interaction contributes 
significantly to improve theory in comparison with experiment. 
This is especially well-demonstrated in the odd mass nuclei for the lowest few excited states. 
The case of the ground state spin of $^{10}$B and its sensitivity to the presence 
of the NNN interaction discussed also in Fig.~\ref{B10_NN}.
is clearly evident. We note that the $^{10}$B results shown in Fig.~\ref{B10_NN}
were obtained with the HO frequency of $\hbar\Omega=15$ MeV, while those
in Fig.~\ref{B10B11C12C13} with $\hbar\Omega=14$ MeV. A weak HO frequency dependence
of the $N_{\rm max}=6$ results is apparent. 

Further details on calculations presented in this section were published 
in Ref.~\cite{NGVON07}. These calculations demonstrate that the chiral NNN 
interaction makes substantial contributions 
to improving the spectra and other observables. However, there is 
room for further improvement in comparison with experiment. 
We used a strength of the 2$\pi$-exchange piece of the NNN interaction,
which is consistent with the NN interaction that we employed (i.e. from Ref.~\cite{N3LO}). 
This strength is somewhat uncertain (see e.g. Ref.~\cite{Nogga06}). Therefore, it will be
important to study the sensitivity of our results 
with respect to this strength.
Further on, it will be interesting to incorporate sub-leading NNN interactions and also
four-nucleon interactions, which are also order N$^3$LO \cite{Epelbaum06}.
Finally, it will be useful
to extend the basis spaces to  $N_{\rm max}=8$ ($8\hbar\Omega$) for $A>6$ 
to further improve convergence.

\section{Beyond nuclear structure with chiral EFT interactions}

\begin{figure}[htb]
\begin{center}
  {\includegraphics[width=0.7\columnwidth]{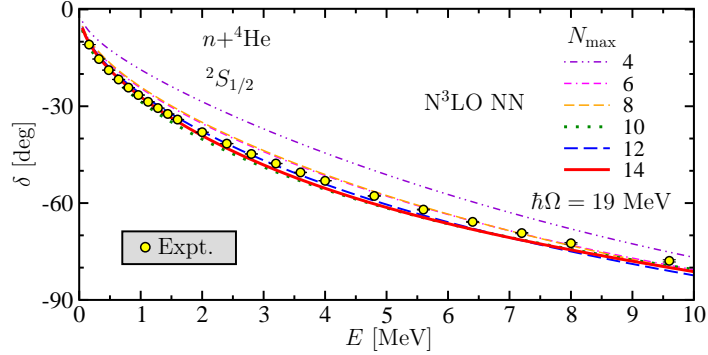}}
  \caption{\label{n_4He_phase_shift}
The calculated $^{2}S_{1/2}$ phase shift for the n+$^4$He system compared with
experimental data. The $\chi$EFT NN potential of Ref.~\cite{N3LO} was used
in model spaces up to the $14\hbar\Omega$.
}
\end{center}
\end{figure}

A shortcoming of the {\it ab initio} NCSM is its incorrect description 
of long-range correlations and its lack of coupling to continuum due to the expansion
of the eigenstates in a finite HO basis. If we want to build upon the {\it ab initio} 
NCSM to microscopically describe nuclear reactions we can proceed in two ways. First,
we can rely on techniques such as the Lorentz Integral Transform that reduce
the continuum problem to a bound-state-like problem. Using this approach, we performed
$^4$He photo-disintegration cross section calculations using the $\chi$EFT 
interactions~\cite{Quaglioni:2007}. We demonstrated a sizeable effect of the chiral NNN
interaction on the cross section. Second, we can augment the {\it ab initio} NCSM basis
by explicitly including cluster states 
and solve for their relative motion while imposing the proper boundary conditions.
This approach, applicable to a wide range of reactions as well as to weakly bound states, 
is in the spirit of the resonating group method (RGM) \cite{RGM}. 
In our approach, we use
the {\it ab initio} NCSM wave functions for the clusters involved and 
effective interactions derived from realistic NN (and eventually also from NNN) potentials.
As an example, a converged calculation of the n+$^4$He $^{2}S_{1/2}$ phase shift 
using the $\chi$EFT NN interaction is presented in Fig.~\ref{n_4He_phase_shift}.
More details on this approach are given in Ref.~\cite{QN_efb20}. 

\section{Conclusions}

The {\it ab initio} NCSM evolved into a powerful many-body technique. Presently,
it is the only method capable to use interactions derived within the $\chi$EFT
for systems of more than four nucleons. 
Among its successes is the demonstration of the importance of the NNN interaction
for nuclear structure. Applications to nuclear reactions with a proper treatment
of long-range properties are under development.
Extension to heavier nuclei is achieved through 
the importance-truncated NCSM \cite{Imp_tr_NCSM}.
Within this approach, {\it ab initio} calculations for nuclei as heavy as $^{40}$Ca 
become possible.

%
\begin{acknowledge}
This work performed under the auspices of the U.S. Department 
of Energy by Lawrence Livermore National Laboratory under Contract DE-AC52-07NA27344.
Support from 
U.S. DOE/SC/NP (Work Proposal Number SCW0498) and the Department of Energy under
Grant DE-FC02-07ER41457 is acknowledged.
\end{acknowledge}

\end{document}